\def\jcap{JCAP}
\def\mnras{MNRAS}             
\def\aap{A\&A}             
\def\apjs{ApJS}             
\def\apjl{ApJL}             
\def\baas{BAAS}

\def\pasj{PASJ}

\documentclass[%
reprint,
superscriptaddress,
nofootinbib,
nolongbibliography,
 amsmath,amssymb,
 aps,
prd,
floatfix,
]{revtex4-2}

\usepackage{graphicx}
\usepackage{dcolumn}
\usepackage{bm}
\usepackage{xcolor}

\definecolor{maroon}{RGB}{192,0,0}
\usepackage[colorlinks=true, citecolor=maroon,
linkcolor=maroon,
urlcolor=blue]{hyperref}
\usepackage{times}

\newcommand{\diffop}{\mathrm{d}}

\newcommand{\mmat}[1]{{\mathbf{#1}}}

\newcommand{\mvec}[1]{{\bm{#1}}}

\newcommand{\cyan}[1]{{\color{cyan} #1}}

\begin{document}

\title{Diffusion-based mass map reconstruction from weak lensing data}

\author{Supranta S. Boruah}
 \email{supranta@sas.upenn.edu} 
\author{Michael Jacob}
 \email{mgjacob@sas.upenn.edu} 
  \author{Bhuvnesh Jain}
 \email{bjain@sas.upenn.edu}
\affiliation{Department of Physics and Astronomy, University of Pennsylvania, Philadelphia, PA 19104, USA}

\date{\today}

\begin{abstract}
Diffusion models have been used in cosmological applications as a generative model for fast simulations and to reconstruct underlying cosmological fields or astrophysical images from noisy data. These two tasks are often treated as separate: diffusion models trained for one purpose do not generalize to perform the other task. In this paper, we develop a  single diffusion model that can be used for both  tasks. By using the Diffusion Posterior Sampling (DPS) approach, we  use a diffusion model trained to simulate weak lensing maps for the inverse problem of reconstructing mass maps from noisy weak lensing data. We find that the standard DPS method leads to biased inference but we  correct this bias by down weighting the likelihood term at early sampling time steps of the diffusion. 
Our method give us a way to reconstruct accurate high-resolution (sub-arcminute) mass maps that have the correct power spectrum and a range of non-Gaussian summary statistics. 
We discuss several applications enabled by the computational efficiency and accuracy of our model. These include generation of simulation quality mass maps, aiding covariance estimation for higher order statistics, and for finding filaments, voids and clusters from noisy lensing shear data.
\end{abstract}

\maketitle


\section{Introduction}\label{sec:intro}

With the advent of Stage IV cosmological surveys such as Euclid \cite{Euclid2011}, Vera Rubin Observatory's Legacy Survey of Space and Time \cite[LSST, ][]{LSST2019}and Roman Space Telescope \cite{Roman_space_telescope}, we are at the beginning of an unprecedented influx of high-quality data. Optimal analysis of these data sets will require major improvements in our data analysis methods. 
Weak gravitational lensing gives us a unique way to probe the underlying matter distribution without relying on biased tracers of dark matter. An important data product for weak lensing analyses are the reconstructions of the underlying mass maps from noisy shear data \cite{DES_Y3_massmapping, HSC_Y1_mass_mapping}. These mass maps are used for many astrophysical applications such as cluster detection \cite{Oguri2024, Chiu2024} and studying the connection of gas and galaxies to the underlying matter. Apart from its astrophysical uses, they are also important for extracting the non-Gaussian cosmological information. 

The late time density field is highly non-Gaussian due to gravitational evolution. But traditional cosmological analyses have relied on 2-point statistics such as correlation functions and power spectra \cite{DESY3_3x2pt_paper, eBOSS_cosmo_2021}, as a result leaving out a large part of the  cosmological information \cite{Boruah2024a, Zurcher2021, Euclid2023}. 
Several studies have attempted to extract the non-Gaussian cosmological information. One  approach uses non-Gaussian summary statistics such as moments of the mass maps \cite{Gatti2022_moments}, wavelet phase harmonics \cite{Gatti2024_WPH}, peak and void counts \cite{Liu2015, Marques2023}, scattering/wavelet transforms \cite{Cheng2020, Ajani2021, Valogiannis2022} or 1-pt PDF \cite{Boyle2021, Thiele2023}, $k$-nearest neighbors \citep[$k$NN,][]{Banerjee2021}. These  summary statistics are usually too complicated to be modelled analytically and therefore we need to rely on simulations for such analyses \cite[e.g,][]{Gatti2024_WPH}. 

Recently the alternate approach of field-level inference (FLI) has been developed with the goal of optimally extracting non-Gaussian information from cosmological data \cite{Jasche2019, Fiedorowicz2022, Boruah2022, Porqueres2022, Dai2024, Zhou2024, Millea2020, Millea2021, Boruah2022_PV, Bayer2023, PrideauxGhee2023}. Inference based on deep learning, primarily based on Convolutional Neural Nets (CNNs), also serve as FLI approaches \cite{Fluri2022, Zhong2024_CNN, Sharma2024, Jeffrey2025}. For FLI as well, we need a field-level generative model for accurate inference. But in addition, we need to solve the inverse problem of reconstructing the cosmological field from noisy data. 
While in principle, we can use $N$-body simulations as our generative model, in practice  computational cost is prohibitive. This necessitates the development of fast and accurate cosmological simulations.   

Generative machine learning methods have been proposed as a fast alternative to $N$-body simulations for producing cosmological fields. Generative models are machine learning methods that are trained to produce samples from the probability distribution of the training data. By training generative models on cosmological simulations, they can produce realistic fields at a fraction of the cost of $N$-body simulations. Many different generative models have been used in cosmological studies. For example, Generative Adversarial Networks (GANs) have been used as fast, surrogate models for $N$-body simulations \cite{Rodriguez2018, KodiRamanah2020, Li2021} as well as for weak lensing and projected density maps \cite{Mustafa2019, Shirasaki2023, Yiu2022, Boruah2024_gansky, Piras2023}. Similarly, normalizing flows have also been used to produce fast cosmological maps \cite{Dai2022, Dai2024, Rouhiainen2021}. 

In this paper, we focus on diffusion-based generative models \cite{Ho2020, Song2020}. In a diffusion model, noise is progressively added to the training data consisting of simulated maps (or images) until the resulting maps are simply maps of random Gaussian noise. Then a neural network is trained to learn (in a probabilistic sense) the reverse `denoising' process. Once trained, the neural network can therefore produce realistic maps from noise. The reverse process requires the neural network to learn the gradient of the log-probability, $\nabla_{\mvec{x}} \log p(\mvec{x})$, which is called the score. Therefore, diffusion models are also  referred to as score-based generative models. 

Diffusion models have also been used in many cosmological and astrophysical applications \cite{Mudur2022, Rouhiainen2024, Ono2024, Legin2024, Floss2024, HeurtelDepeiges2023, Remy2023, Rouhiainen2024, CuestaLazaro2024, Mudur2024}. In astrophysics, diffusion models have been used for two tasks: First, as a generative models for producing fast simulations. For example, \cite{Mudur2022} used diffusion models as a generative model for dust maps and dark matter simulations and  \cite{Rouhiainen2024} used them to produce super-resolution simulations. Second, diffusion models are used to reconstruct an underlying astrophysical field from noisy data 
\cite{Ono2024, Legin2024, Floss2024, HeurtelDepeiges2023}. In the first case, the diffusion model is trained on simulations without conditioning it on data while in the second case the diffusion model is trained conditioned on noisy data realizations and cannot be used to generate the  underlying (noiseless) astrophysical fields. However, we are often able to forward model the noisy data realization given an astrophysical field. This suggests that we may be able to reconstruct astrophysical fields/images using an unconditioned diffusion model by forward modeling the noisy data likelihood during the sampling step. That is, we can attempt to use the same diffusion model to produce simulation realizations as well as for the inverse problem of reconstructing the underlying field given a noisy data realization.
However, solving this inverse problem with a diffusion model inevitably requires approximating the likelihood with respect to the latent variables of the diffusion model, since the exact solution is intractable. Such a unified approach has been used for astrophysical image reconstruction \cite[e.g,][]{Adam2023, Dia2023}, and for reconstructing mass maps from noisy shear data \cite{Remy2023}.

In this paper we use the Diffusion Posterior Sampling \cite[DPS, ][]{Chung2022_DPS} approach to solve the inverse problem of reconstructing mass maps from noisy weak lensing shear data.
In this approach, we train a diffusion model without conditioning it on the data and therefore use it as a `prior' for the underlying field. To reconstruct the underlying field given noisy data requires us to evaluate the gradient of the likelihood, $\nabla_{\mvec{x}} \log p(\mvec{d}|\mvec{x})$. As mentioned, this gradient term is not tractable since it requires marginalizing over all latent variables of the diffusion model. However, the method of \cite{Chung2022_DPS} allows us to approximate this likelihood gradient term. 

We validate our model by testing it both as a generative model and its mass map reconstruction. Our generative model reproduces both Gaussian as well as non-Gaussian statistics of the convergence maps from $N$-body simulations. We further show that the weak lensing mass maps reconstructed using the DPS approach are accurate to sub-arcminute scales, making it one of the few algorithms that can perform accurate, high-resolution reconstruction.

There are multiple benefits to using our diffusion-based approach. First, we can use the same diffusion model as a generative model as well as for reconstructing cosmological fields allowing us to isolate and test different parts of the generative and reconstruction process. Next, since we do not need to condition our model on noisy data for training our diffusion model, it greatly reduces the complexity of the model, i.e. the diffusion model does not need to learn observational details such as the effect of the mask. Finally, Hamiltonian Monte Carlo (HMC)-based field inference methods  potentially suffer from problems such as long auto-correlation lengths \cite{Millea2020} and multi-modality of the high-dimensional posteriors \cite{Feng2018}. The DPS approaches can mitigate both these problems since each sample created from the diffusion model is independent, which makes it an attractive option for Bayesian field-level inference in the future.

This paper is structured as follows: we describe the weak lensing theory and the simulations used in this work in section \ref{sec:wl} and the diffusion model training and the posterior inference scheme in section \ref{sec:methodology}. We present our results in section \ref{sec:results} before concluding in section \ref{sec:conclusion}. 
\section{Weak lensing theory and simulations}\label{sec:wl}

\subsection{Weak lensing basics}\label{ssec:wl_basics}

The large-scale matter distribution in the Universe slightly distorts the shapes of galaxies through weak gravitational lensing. The magnitude of the galaxy shape distortion is related to the shear field, $\gamma$, that is in turn determined by the convergence field, $\kappa$.  The convergence field is given in terms of the matter overdensity, $\delta$, as
\begin{equation}
    \kappa(\mvec{\theta}) = \int_{0}^{\chi_{\text{H}}} d \chi\ g(\chi) \delta(\chi \mvec{\theta}, \chi),
\end{equation}
where, $g(\chi)$ is the lensing efficiency
\begin{equation}
    g(\chi) = \frac{3 H^2_0 \Omega_m}{2 c^2}\frac{\chi}{a(\chi)} \int_{\chi}^{\chi_{\text{H}}} d \chi^{\prime} \frac{\chi^{\prime} - \chi}{\chi^{\prime}} p(\chi^{\prime}).
\end{equation}
Here, $p(\chi)$ is the redshift probability distribution of the source galaxies, $\chi$ denotes the comoving distance, $\chi_{\text{H}}$ is the comoving horizon distance. The relationship between the shear field and the convergence field is given by the Kaiser-Squires (KS) transform \cite{Kaiser1993}. The KS transform relates the Fourier transform of the convergence field to the Fourier transform of the two components of the shear fields via
\begin{align}\label{eqn:KS1}
    \tilde{\gamma}^{1}_{\mvec{l}} &= \frac{l^2_x - l^2_y}{l^2} \tilde{\kappa}_{\mvec{l}},\\
    \tilde{\gamma}^{2}_{\mvec{l}} &= \frac{2 l_x l_y}{l^2}\tilde{\kappa}_{\mvec{l}}. \label{eqn:KS2}
\end{align}
Thoughout this paper, we assume a spatially flat Universe and the flat-sky approximation. In our notation, we denote our Fourier modes with $\tilde{[.]}$.  

The observed galaxy shapes are a noisy realization of the underlying shear field. In this work we assume that the shape noise in each pixel is independent of other pixels.\footnote{This assumption no longer holds in the presence of intrinsic alignment.} With this assumption, the likelihood of observing a noisy shear map, $\gamma_{\text{obs}}$ is given as, 
\begin{equation}\label{eqn:shear_likelihood}
    P(\gamma_{\text{obs}}|\kappa) \propto \prod_{\alpha=1}^2 \prod_{i=1}^{N_{\text{pix}}} \exp \bigg[-\frac{(\gamma^i_{\text{obs},\alpha} - \gamma^i_{\alpha}(\kappa))^2}{2\sigma^2_{\epsilon, i}}\bigg]         
\end{equation}
where, $\alpha$ denotes the two shear components, $\sigma_{\epsilon,i}$ denotes the shape noise in the $i$-th pixel, and $\gamma_{\alpha}(\kappa)$ is given by equations \eqref{eqn:KS1} and \eqref{eqn:KS2}. 

\subsection{Mass mapping with weak lensing}\label{ssec:wl_massmapping}
\subsubsection{Gaussian prior}

Since weak lensing observations probe the underlying matter distribution, we can use these observations to create maps of matter distribution in the Universe, including the contribution of dark matter that is otherwise not directly observed. Perhaps the most popular approach for mass map reconstruction from weak lensing is the Kaiser-Squires inversion \cite{Kaiser1993}. Another popular approach uses Wiener filtering to suppress the small scale noise in the mass maps. Both these approaches can be interpreted in a single unified Bayesian framework where the reconstructed mass maps are created by either optimizing or sampling from a posterior,
\begin{equation}\label{eqn:kappa_posterior}
    P(\mvec{\kappa}|\mvec{d}) \propto P(\mvec{d}|\mvec{\kappa})P(\mvec{\kappa}),
\end{equation}
where, $P(\mvec{d}|\mvec{\kappa})$ is the likelihood given by the equation \eqref{eqn:shear_likelihood} and $P(\mvec{\kappa})$ is the prior on the convergence field \cite{DESY3_massmapping}. In this framework, the KS inverted map is the maximum-likelihood map and the Wiener filtered map is the maximum-a-posteriori map assuming a Gaussian prior on the convergence field. Rather than creating one best estimate of the mass map, other approaches \cite{Alsing2016, Alsing2017, Kovacs2022, Loureiro2023, Sellentin2023} have tried to create a sample of mass maps drawn from a Wiener posterior. Samples of mass maps from the posterior allows for better uncertainty quantification that is not possible with a single best estimate of KS or Wiener filtered maps.

\subsubsection{Non-Gaussian mass map reconstruction}\label{ssec:ng_wl}

Although the initial conditions of the Universe are well-described as Gaussian, the gravitational dynamics make the late-time convergence field highly non-Gaussian. Thus, for correctly describing the late-time convergence field, we need to go beyond using a Gaussian prior in equation \eqref{eqn:kappa_posterior}. One simple analytic extension assumes that the convergence field can be described as a lognormal random field \cite{Xavier2016}. Indeed, lognormal priors have been used in the Bayesian mass map reconstruction framework to create weak lensing mass maps \cite{Fiedorowicz2022, Fiedorowicz2022a, Boruah2022, Zhou2024, Boruah2024_massmapping}. But in spite of its successes, the lognormal model is insufficient to describe the convergence field at the accuracy requirements of Stage-IV surveys \cite{Boruah2024_gansky}. 
We note that analytical extensions with Generalized Point Transformed Gaussian (GPTG) transforms \cite{Zhong2024} might be successful in modeling the convergence field at these accuracy levels.

Generative machine learning approaches have emerged as a popular alternative to modeling the convergence field. For example, \cite{Mustafa2019, Shirasaki2023, Yiu2022, Boruah2024_gansky} used GANs to model the late-time convergence field. Similarly, \cite{Dai2022, Dai2024} used normalizing flows to model the convergence field. Once a generative model is trained, it can be used to get the prior in equation \eqref{eqn:kappa_posterior}.

In this paper, we use diffusion models to learn the distribution of convergence maps. In diffusion models, a neural network is trained to learn the gradient of the log-prior, $\nabla_{\mvec{\kappa}} \log P(\mvec{\kappa})$ or the score function. We describe diffusion models in detail in section \ref{sec:methodology}.

\subsection{Weak lensing simulations}\label{ssec:wl_sims}

We train our diffusion model using the publicly available weak lensing maps from the \texttt{MassiveNuS} simulation suite \cite{MassiveNu2018}.\footnote{\url{https://columbialensing.github.io/\#massivenus}} The full simulation suite contains 101 sets of simulations with varying cosmological parameters. For this work, we only use the fiducial simulation with massless neutrinos and cosmological parameter values of $\Omega_m = 0.3$ and $\sigma_8 = 0.8523$. Each simulation set contains 10,000 convergence maps created by ray-tracing the $N$-body simulations. These maps are provided at five different redshifts: $z = 0.5, 1.0, 1.5, 2.0,$ and $2.5$. For our model, we use maps at three redshifts: $z = 0.5, 1.0,$ and $1.5$. The choice of three bins (and not more) is done for simplicity, with the redshifts chosen to span the range with most of the information from a Stage 4 survey. Each convergence map covers an area of $(3.5 \text{ deg})^2$ and is gridded into $1024^2$ pixels. We regrid these maps into $256 \times 256$ pixels, corresponding to a pixel resolution of $0.8$ arcmin. In total, we use 8,000 convergence maps to train our diffusion model.

\subsubsection{Creating noisy mock data}\label{sssec:noisy_mock_data}

In section \ref{ssec:dps_mass_mapping}, we infer the underlying convergence maps from mock noisy observations. To do so, we first compute the shear field from the simulated convergence maps using equations \eqref{eqn:KS1} and \eqref{eqn:KS2}. We then create a noisy mock shear data by adding a Gaussian shape noise in each pixel with variance given by, 
\begin{equation}
    \sigma^2_{\epsilon,\text{pix}} = \frac{\sigma^2_{\epsilon}}{A_{\text{pix}}n_{\text{eff}}},
\end{equation}
where, $A_{\text{pix}}$ is the pixel area, $n_{\text{eff}}$ is the effective number density and $\sigma_{\epsilon}$ is the shape noise per ellipticity component per galaxy. We assume $\sigma_{\epsilon}=0.26$ and $n_{\text{eff}}=7.5$ arcmin$^{-2}$ in each of the three tomographic bins. This roughly corresponds to the source number density expected in the LSST-Y10 survey (though the LSST analysis will very likely be done with more tomographic bins).

\section{Methodology}\label{sec:methodology}

\subsection{Basics of diffusion models}

In diffusion models, we progressively add noise to create a series of noisy images, $\mvec{x}(t)$ labelled by a `time' variable, $t$. The progression of the images through time defines a diffusion process. This `forward' diffusion process is modelled as a stochastic differential equation (SDE), 
\begin{equation}\label{eqn:fwd_diffusion}
    \diffop\mvec{x} = f(\mvec{x}, t) \diffop t + g(t) \diffop \mvec{w} , 
\end{equation}
where, $\diffop \mvec{w}$ is standard Wiener noise and $f$ and $g$ are called the drift and the diffusion coefficient respectively. Different functions for $f$ and $g$ defines different diffusion processes. Given the forward diffusion process in equation \eqref{eqn:fwd_diffusion}, its reverse process can also be modelled as another SDE \cite{Anderson1982}, 
\begin{equation}\label{eqn:reverse_diffusion}
    \diffop\mvec{x} = [f(\mvec{x}, t)  - g^2(t)\nabla_{\mvec{x}}\log p_t(\mvec{x})] \diffop t + g(t) \diffop\mvec{w},
\end{equation}
where, $p_t(\mvec{x})$ is the probability distribution of the images (or maps) at time $t$. Thus, we can reverse the `noising' diffusion process (in a probabilistic sense) using the gradient of log-probability (or the `score' function). This is the key result on which diffusion models are based. In these models, noise is progressively added to the training images (or maps) following equation \eqref{eqn:fwd_diffusion} so that it ultimately results in pure Gaussian random noise. 

We then need the score to reverse this diffusion process. In deep-generative score-based models, a neural network is used to learn the score at different timesteps, so that we can approximate the score function as,
\begin{equation}
    \nabla_{\mvec{x}} \log p_t(\mvec{x}) \approx \mvec{s}_{\mvec{\theta}}(\mvec{x}, t),
\end{equation}
where, $\mvec{s}$ is the score neural network and  $\mvec{\theta}$ are the parameters of the neural network that are fitted during training. Once the score network is trained, we can use the score in equation \eqref{eqn:reverse_diffusion} to reverse the diffusion process and sample an image (or a map) from the target distribution. 

In practice, both the forward and reverse diffusion SDEs are discretized into a set number of timesteps, $T$. The `forward' diffusion model is chosen such that the probability distribution at time $T$ is pure Gaussian noise with mean $\mvec{\mu}_T$ and standard deviation, $\sigma_T$, i.e,
\begin{equation}
    \mvec{x}_T \sim  \mathcal{N}(\mvec{x}; \mvec{\mu}_T, \sigma^2_T).
\end{equation}
In this paper, we train our diffusion model using the Denoising Diffusion Probabilistic Model (DDPM) approach \cite{Ho2020}, where the forward diffusion is modelled as,
\begin{equation}\label{eqn:ddpm_fwd}
    q(\mvec{x}_t|\mvec{x}_{t-1}) = \mathcal{N}(\mvec{x}; \sqrt{1-\beta_t} \mvec{x}_{t-1}; \beta_t \mmat{I}).
\end{equation}
The coefficient, $\beta_t$ is the variance of the Gaussian noise added at timestep $t$ and it is chosen to be between $0$ and $1$ in an increasing sequence in time. Equation \eqref{eqn:ddpm_fwd} corresponds to a discretized version of equation \eqref{eqn:fwd_diffusion} with $f(t)=-\beta(t)\mvec{x} / 2$, and $g(t)=\sqrt{\beta(t)}$ \cite{Song2020}. We discretize our diffusion process in $1000$ timesteps and chose a sigmoid variance schedule for $\beta_t$ \cite{Jabri2022}.

In DDPM, the reverse step probability is parameterized as a Gaussian, 
\begin{equation}
    p_{\theta}(\mvec{x}_{t-1}|\mvec{x}_{t}) = \mathcal{N}(\mvec{x}_{t-1}; \mvec{\mu}_{\mvec{\theta}}(\mvec{x}_t, t), \sigma^2_t \mmat{I}),
\end{equation}
where,
\begin{equation}       
    \mvec{\mu}_{\mvec{\theta}}(\mvec{x}_t) = \frac{1}{\sqrt{1-\beta_t}}\bigg[ \mvec{x}_t - \frac{\beta_t}{\sqrt{1 - \bar{\alpha}_t}}\mvec{\epsilon}_{\mvec{\theta}}(\mvec{x}_t, t)\bigg],   
\end{equation}
and $\sigma^2_t = \beta_t$. Here, $\mvec{\epsilon}_{\mvec{\theta}}(\mvec{x}_t, t)$ is predicted using the score neural network. In the above equation, $\bar{\alpha}_t = \prod_{s=1}^{t}(1-\beta_s)$. The score network is then trained by minimizing the loss function, 
\begin{equation}\label{eqn:ddpm_loss}
    L(\mvec{\theta}) = \underset{\mvec{\epsilon}, \mvec{x}_0}{\mathbb{E}} \bigg[ \lVert \mvec{\epsilon} - \mvec{\epsilon}_{\mvec{\theta}}(\sqrt{\bar{\alpha}_t}\mvec{x}_0 + \sqrt{1 - \bar{\alpha}_t\mvec{\epsilon}}, t) \rVert ^2\bigg],
\end{equation}
with respect to the neural network parameters $\mvec{\theta}$. In the above equation, $\mvec{\epsilon} \sim \mathcal{N}(0, \mmat{I})$ and $\mvec{x}_0$ are the set of images from the training dataset. This loss function is related to the KL divergence of the two Gaussian distributions of the forward and the reverse process. 
In practice, a new set of Gaussian noise, $\mvec{\epsilon}$, is resampled at each training step and the loss in equation \eqref{eqn:ddpm_loss} is computed with  randomly chosen noise map, $\mvec{\epsilon}$.
For more details, we refer the reader to \cite{Ho2020}.

We use a UNet architecture \cite{UNet2015} for our score network. There are $4$ downsampling layers, the first consisting of 64 feature channels and up to 512 feature channels in the last layer. Each downsampling layer consists of 2 ResNet blocks \cite{ResNet2016}, followed by an application of a multi-head linear attention layer \cite{Vaswani2017}, and finally a downsampling operation. Upsampling layers are structured the same way, with 2 ResNet blocks, an attention layer, and an upsampling operation, but also with features from the corresponding downsampling block as input. Since the score network needs to predict the denoiser as a function of time, a multilayer perceptron is used to create a sinusoidal time embedding that is fed into each of the ResNet blocks. In total our network consists of 35 million trainable parameters. We train the network for 300000 iterations. This UNet architecture  structure and the diffusion model structure we use are both adapted from the Denoising-Diffusion-PyTorch repository\footnote{\url{https://github.com/lucidrains/denoising-diffusion-pytorch}}, which implements the framework introduced by \cite{Ho2020}.  

\subsubsection{Data pre-processing with analytical transforms}

While training our diffusion model, we found that pre-processing the $\kappa$ maps with certain analytical transforms helps in better performance of the model. Here we describe these pre-processing steps. It is well-known that the convergence field can be well represented as a lognormal field \cite{Taruya2002,Clerkin2017, Boruah2022, Zhou2024}, 
\begin{equation}\label{eqn:LN_transform}
    \kappa = e^y - \kappa_{\text{min}},
\end{equation}
where, $y$ is a Gaussian random field. Therefore, we first use the inverse transform of equation \eqref{eqn:LN_transform} to get the $y$ maps from the $\kappa$ maps. The value of $\kappa_{\text{min}}$ is determined from the simulations. Next, we normalize the $y$ maps to have values between $0$ and $1$. 
\begin{equation}\label{eqn:y_norm}
    \tilde{y} = \frac{y - y_{\text{min}}}{y_{\text{max}} - y_{\text{min}}},
\end{equation}
where, $y_{\text{min/max}}$ are the minimum/maximum values of $y$. We train our diffusion model to predict $\tilde{y}$ from which $\kappa$ is determined using equation \eqref{eqn:LN_transform} and \eqref{eqn:y_norm}. Note that these transforms are applied independently on each of the tomographic bins.

\subsection{Posterior sampling with the diffusion model}\label{ssec:dps_methodology}

Solving an inverse problem, e.g, reconstructing the mass map from noisy weak lensing data involves sampling from a posterior distribution, $p(\mvec{x}|\mvec{d})$. Using diffusion models for this purpose requires replacing the score term in equation \eqref{eqn:reverse_diffusion} with the gradient of the posterior, i.e, $\nabla \log p(\mvec{x}|\mvec{d})$. Using Bayes' Theorem we can write the gradient of the log-posterior as, 
\begin{align}\label{eqn:posterior_diffusion}
    \nabla_{\mvec{x}_t}\log p_t(\mvec{x}_t|\mvec{d}) &= \nabla_{\mvec{x}_t}\log p(\mvec{d}|\mvec{x}_t) + \nabla_{\mvec{x}_t} \log p_t(\mvec{x}_t) \nonumber \\
    &\approx \nabla_{\mvec{x}_t} \log p(\mvec{d}|\mvec{x}_t) + \mvec{s}_{\mvec{\theta}}(\mvec{x}_t, t)
\end{align}
However, evaluating the likelihood term in the above equation is non-trivial and requires marginalizing over all possible trajectories that results in $\mvec{x}_t$ at time $t$, 
\begin{align} \label{eqn:intractable_lkl}
    p(\mvec{d}|\mvec{x}_t) &= \int d\mvec{x}_0 p(\mvec{d}|\mvec{x}_0) p(\mvec{x}_0|\mvec{x}_t) \nonumber \\
    &= \mathbb{E}_{\mvec{x}_0 \sim p(\mvec{x}_0|\mvec{x}_t)}[p(\mvec{d}|\mvec{x}_0)]. 
\end{align}
Thus, evaluating equation \eqref{eqn:intractable_lkl} is not analytically tractable showing the difficulty of using diffusion models for solving inverse problems. \cite{Chung2022_DPS} introduced an approximate Diffusion Posterior Sampling (DPS) approach to approximate the likelihood term to enable solving inverse problems in diffusion models. In their method, the likelihood term is approximated by, 
\begin{equation}\label{eqn:jensen_approx}
    p(\mvec{d}|\mvec{x}_t) \approx p(\mvec{d}|\hat{\mvec{x}}_0), 
\end{equation}
where, $\hat{\mvec{x}}_0$ is the posterior mean of $p(\mvec{x}_0|\mvec{x}_t)$ given by, 
\begin{equation}
    \hat{\mvec{x}}_0 = \mathbb{E}[{\mvec{x}_0|\mvec{x}_t}] = \int d\mvec{x}_0 \mvec{x}_0 p(\mvec{x}_0|\mvec{x}_t) .
\end{equation}
In doing so, the expectation value is pulled inside equation \eqref{eqn:intractable_lkl}. The error in this approximation has an upper bound given by  Jensen's inequality. $\hat{\mvec{x}}_0$ is given by Tweedie's formula \cite{Tweedies_formula} and can be evaluated using the score network as, 
\begin{align}\label{eqn:tweedie}
    \hat{\mvec{x}}_0(\mvec{x}_t, t) &= \frac{1}{\bar{\alpha}(t)}\bigg[\mvec{x}_t + (1 - \bar{\alpha}(t))\nabla_{\mvec{x}_t}\log p_{t}(\mvec{x}_t)\bigg] \nonumber \\
    &\approx \frac{1}{\bar{\alpha}(t)}\bigg[\mvec{x}_t + (1 - \bar{\alpha}(t))\mvec{s}_{\mvec{\theta}}(\mvec{x}_{t}, t)\bigg].
\end{align} 
Thus, in order to sample maps from the posterior using the DPS approach, we need to replace the score term in the sampling step to include the gradient of the likelihood term, 
\begin{equation}\label{eqn:dps_score}
    \mvec{s}_{\mvec{\theta}}(\mvec{x}_t) \rightarrow \mvec{s}_{\mvec{\theta}}(\mvec{x}_t) + \nabla_{\mvec{x}_t} \log p(\mvec{d}|\hat{\mvec{x}}_0 (\mvec{x}_t, \mvec{s}_{\mvec{\theta}}(\mvec{x}_t))). 
\end{equation}
For mass map sampling, this likelihood term is given in equation \eqref{eqn:shear_likelihood} in section \ref{sec:wl}. We implement the likelihood in {\sc pytorch} where the gradient of the likelihood can be computed using automatic differentiation.  

\subsubsection{Controlling the bias in the DPS algorithm}

As we will see in the next section, the standard DPS algorithm as implemented in equation \eqref{eqn:dps_score} leads to a bias in the inferred maps. This bias arises because of errors accumulated in the early time of the diffusion sampling and has been noted in the literature \cite[e.g. ][]{sitcom, daps}. The measurement noise and the approximation errors in the DPS algorithm results in latent map samples, $\mvec{x}_t$, from the low-probability regions of the prior, $p_t(\mvec{x}_t)$. The errors are accumulated over the large number of timesteps of the diffusion sampling process resulting in highly biased samples $\mvec{x}_0$. 

In this paper, we introduce a new scheme where we down-weight the likelihood gradient at early times. We implement this downweighting by multiplying the likelihood gradient with a scaling parameter, $A(t)$, with the following constraints: {\it i)} At early times, i.e, $t=t_{\text{end}}$, $A(t) \rightarrow 0$ at, and {\it ii)} At late times, i.e, $t=0$, $A(t) \rightarrow 1$. We use a sigmoid function to model the likelihood scaling factor, \begin{equation}\label{eqn:lkl_rescaling}
    A(t) = \text{Sigmoid}\bigg[\frac{\bar{t}-t}{\sigma_t}\bigg].
\end{equation} 
We determine $\bar{t}, \sigma_t$ empirically and found that $\bar{t}=500$, $\sigma_t=20$ leads to unbiased posterior inference. 

\begin{figure}
    \centering
    \includegraphics[width=\linewidth]{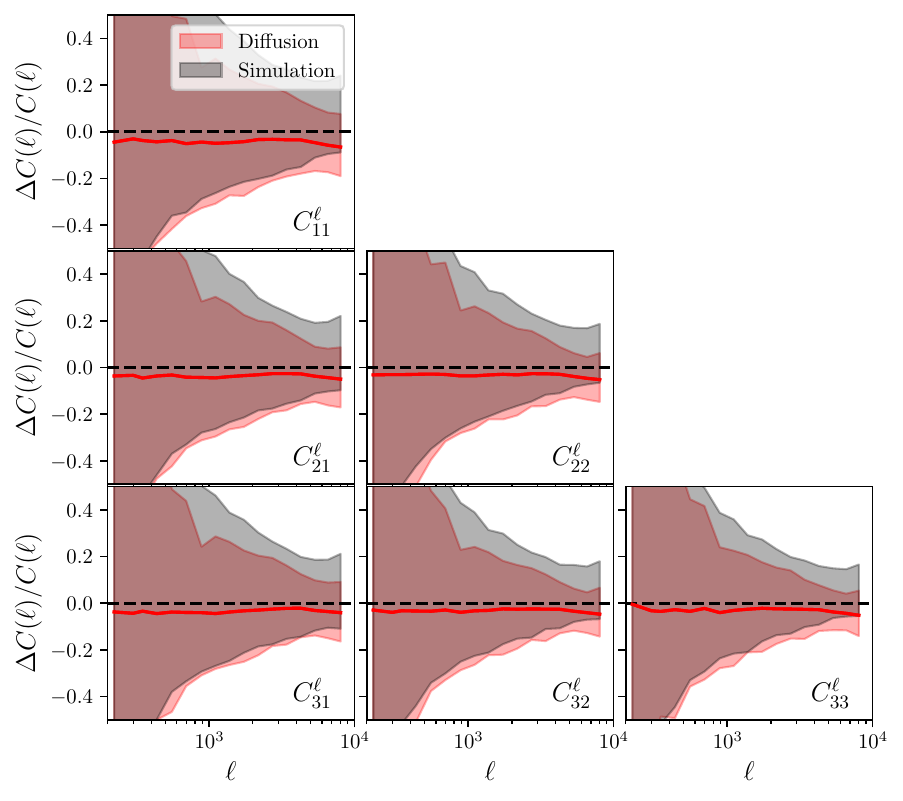}
    \caption{Comparison of the power spectrum of the simulation maps (\textit{black}) and the diffusion maps (\textit{red}). The different panels show the power spectrum between different redshift bins as labeled. The solid line is the median and the shaded region is the $95\%$ confidence region of the power spectrum estimated from 1024 samples. As we can see from the figure, the diffusion model has a bias at  the few percent level in the power spectrum. 
    Note that at  $\ell$ above about $10^3$, shape noise (not included here) will dominate the uncertainty. 
    }
    \label{fig:Cl_plot}
\end{figure}

\begin{figure*}
    \centering
    \includegraphics[width=\linewidth]{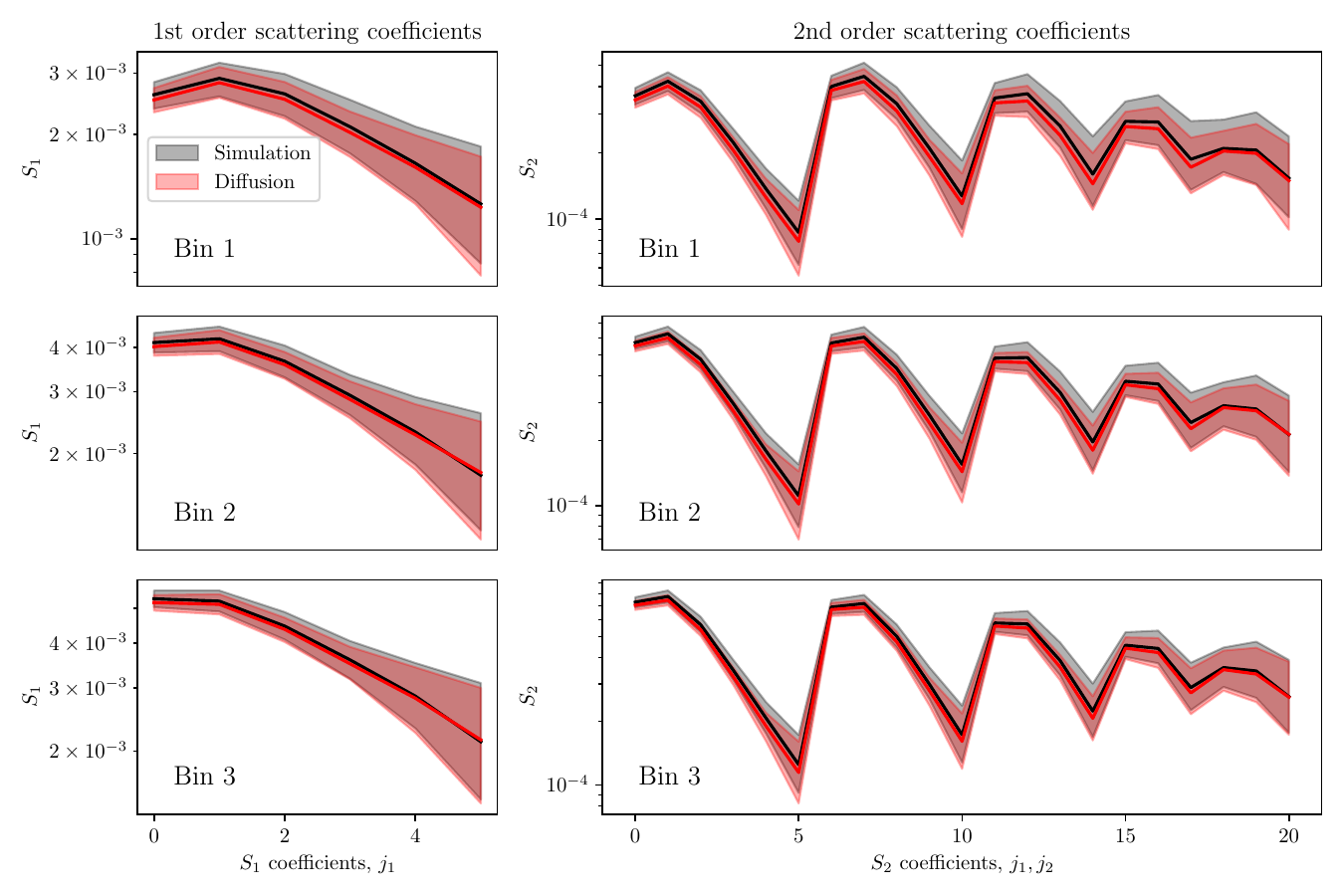}
    \caption{Comparison of scattering transform measured from the simulation maps (\textit{black}) and maps generated by our diffusion model. The left column shows the first order scattering transform and the right column shows the 2nd order scattering transform. The three rows show the statistics for the $3$ redshift bins. Clearly the diffusion model is an accurate generative model. In Figs. \ref{fig:posterior_st} and \ref{fig:posterior_ng_stats} we show that this is also the case for mass map reconstruction. 
    }
    \label{fig:st}
\end{figure*}

\section{Results}\label{sec:results}

\subsection{Diffusion model as a generative model}

We train our diffusion model on the convergence maps described in section \ref{ssec:wl_sims}. Once trained, the diffusion model can generate maps with same statistical properties as the simulations on which the model is trained. However, since no model is perfect, we need to validate that the trained model indeed reproduces the statistics of the simulations maps with reasonable accuracy. To do so, we generate $1024$ convergence maps with our diffusion model and compare various summary statistics of the generated maps with the same number simulation maps from the validation set. 

We first compare the power spectrum of the maps in Figure \ref{fig:Cl_plot}. As we see from the figure, the bias in the power spectrum of the generated maps is less than a few percent. As we can see from the figure, the diffusion maps not only have the correct auto-power spectra, but also reproduces the cross-power spectra of the simulation maps between different redshift bins showing that it correctly models the correlation between different bins.

One of the main advantages of using generative AI methods for WL map generation is that they better can capture the non-Gaussian information of the field that is not well represented by simple analytical models such as the lognormal model. To test how well they do, we use a set of non-Gaussian summary statistics. The wavelet based statistic known as scattering transforms \cite[ST,][]{Cheng2020}   provide a compact way to extract the cosmological information from convergence fields. Therefore, we compare the scattering transform computed from the two sets of maps in Figure \ref{fig:st} (in Appendix A we present other summary statistics to validate mass map reconstruction).  Computing scattering transforms involves first convolving the field with a set of wavelets and then taking its absolute value. Then the scattering coefficients are calculated by taking the average of the convolved field. We can compute the first-order or the second-order scattering transform by performing the field transform once or twice. 
Specifically, we compute  $S_1$ and $S_2$ as a function of the scale coefficients, $j_1/j_2$, as,
\begin{align}
    S_1(j_1) &= \langle |\kappa * \psi(j_1)|\rangle,\\
    S_2(j_1, j_2) &= \langle ||\kappa * \psi(j_1)| * \psi(j_2)\rangle.
\end{align}
Here, $\psi$ denotes the wavelets used for filtering the maps and $j_1 / j_2$ determine the wavelet scale. 
We compute only up to the second order as \cite{Cheng2020} showed that the first two orders are sufficient for extracting cosmological information from convergence fields. We refer the reader to \cite{Cheng2020} for details on ST. As we can see from the figure, the diffusion model correctly reproduces both the first-order and second-order scattering transforms of the convergence maps.

\begin{figure*}
    \centering
    \includegraphics[width=\linewidth]{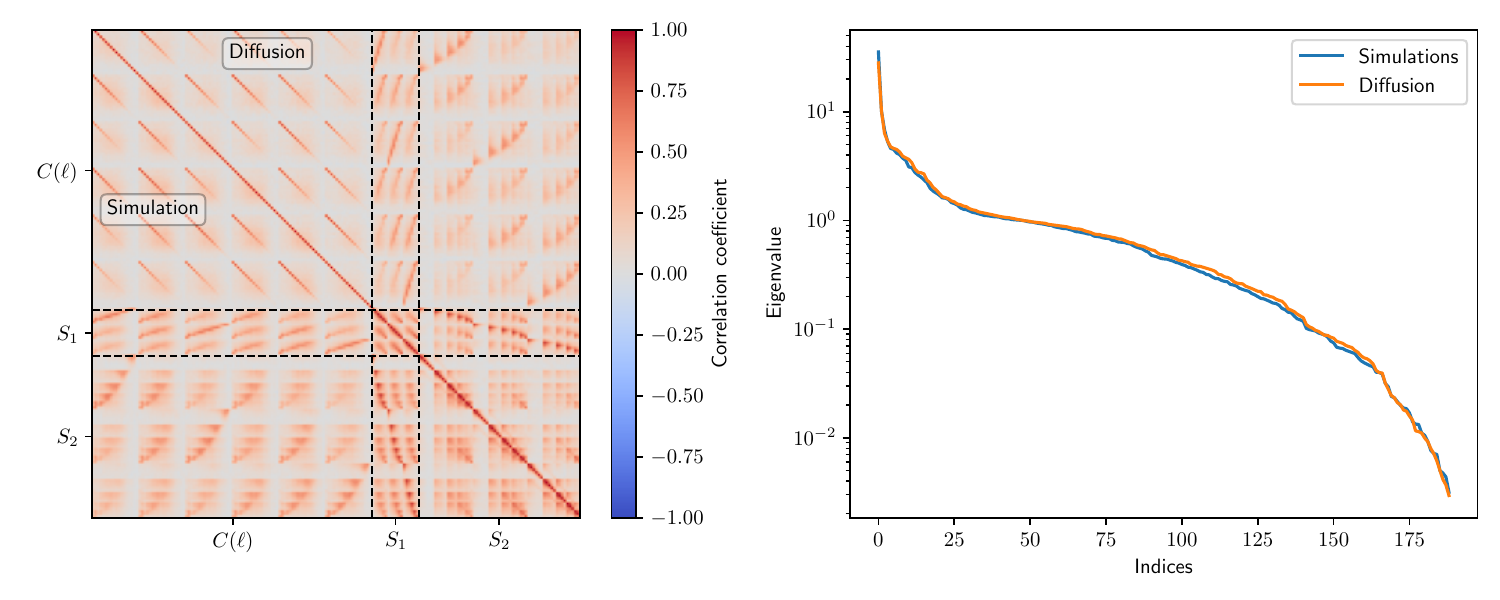}
    \caption{Comparison of the covariance matrix computed from the simulations and the trained diffusion model. The left  panel shows the correlation matrix  -- the lower left matrix is from the simulations, while the upper right shows the same from the diffusion model. The two matrices are in excellent agreement. The right hand panel compares the eigenvalues of the two covariance matrices showing the fidelity of the diffusion-based covariance matrix. 
    }    
    \label{fig:cov}
\end{figure*}

\begin{figure*}
    \centering
    \includegraphics[width=\linewidth]{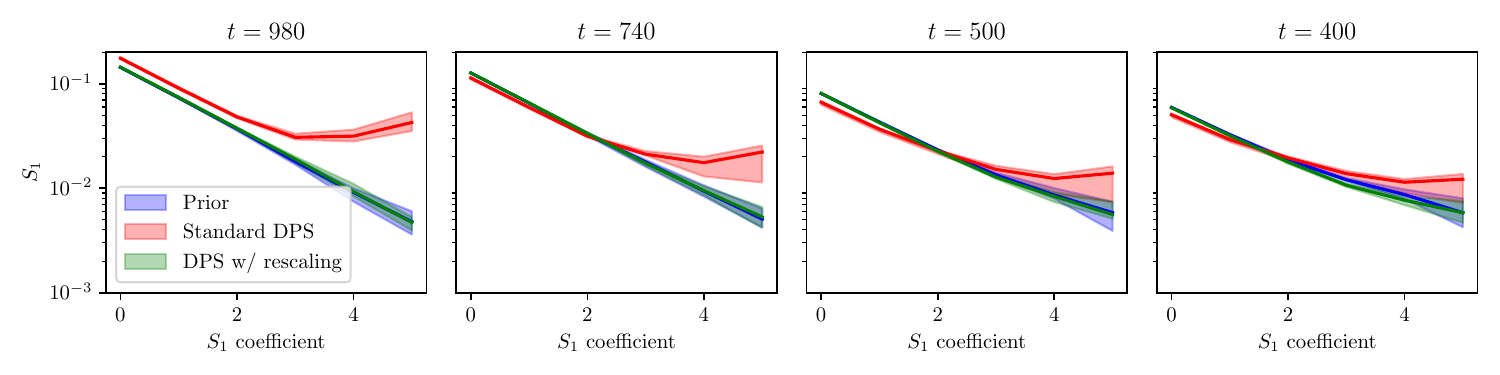}
    \caption{The first order scattering transform of the latent variable maps at different time steps as labelled in the figure. The blue curves show the scattering transforms of latent maps sampled from the prior. As we can see from the figure, in the standard DPS algorithm, the latent maps in the early stages of the sampling go out-of-distribution leading to biased inference at $t=0$. However, the simple likelihood re-scaling scheme that we introduce (shown in green) can mitigate this bias. 
    }
    \label{fig:S1_ood}
\end{figure*}

Figures \ref{fig:Cl_plot}, and \ref{fig:st} show that the diffusion model is largely successful in reconstructing the different summary statistics of the convergence maps. If the diffusion model is indeed successful in capturing the full probability distribution of the convergence fields, then we can use it to also compute the covariance matrix of the various summary statistics we have considered. Covariance matrices are useful for performing cosmological inference assuming a Gaussian likelihood for the summary statistics or to perform data compression for simulation-based inference \cite{Park2024}. To compare the covariance matrices, we create noisy shear maps from the convergence maps created with the diffusion model and the simulation maps. We compute the different summary statistics from these noisy maps and compute the covariance matrix. The correlation matrix and the eigenvalues of the covariance matrix are shown in Figure \ref{fig:cov}: the correlation matrix computed with the diffusion model has the same  structure as  the the one from simulations showing that the diffusion model is able to learn the joint distribution of different non-Gaussian summary statistics. Furthermore, we see that the covariance matrix computed with the diffusion model and that computed from the simulations have similar eigenvalues showing the fidelity of the diffusion-based covariance matrix.

\begin{figure*}
    \centering
    \includegraphics[width=\linewidth]
    {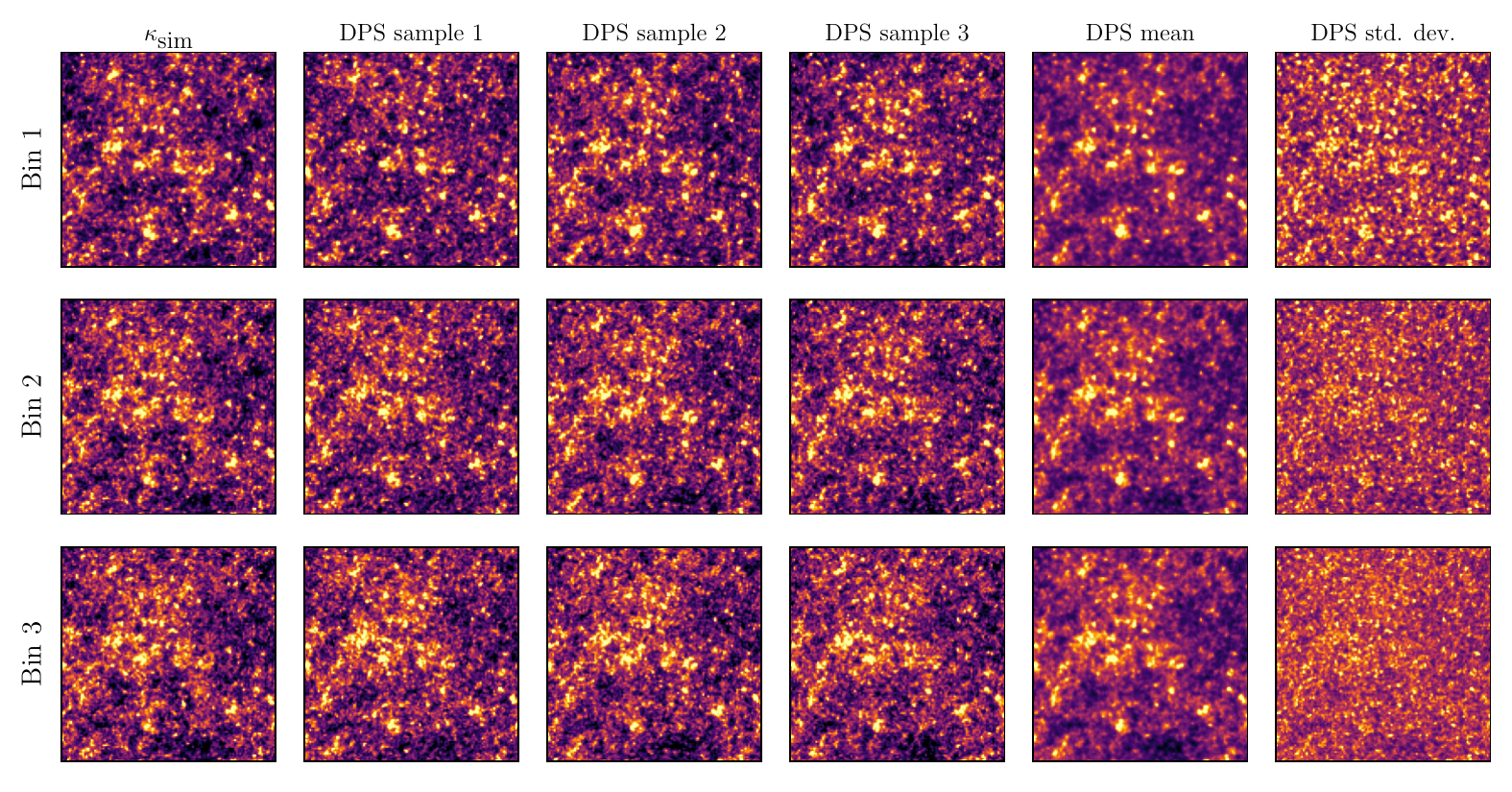}
    \caption{Comparison of the diffusion posterior maps with the true underlying simulated convergence maps (\textit{left}). The second, third and fourth columns show randomly selected sample maps from the posterior. The fourth column shows the mean map computed from 100 sample maps from the diffusion posterior. Finally, the rightmost column shows the uncertainty in the reconstruction measured with the standard deviation in the posterior maps. The three rows correspond to the three redshift bins. }
    \label{fig:map_plot_posterior}
\end{figure*}

\begin{figure}
    \centering
    \includegraphics[width=\linewidth]{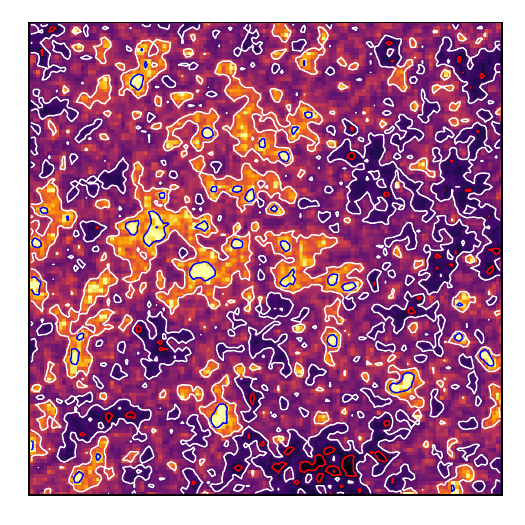}
    \caption{The mean map in the third tomographic bin with the signal-to-noise ratio (SNR) contours over-plotted. The white contours show the SNR$=1$ and the blue (red) contours shows SNR$=3$ in the overdense (underdense) region. As we can see from the figure, high SNR peaks and voids are robustly detected in the mass map.}
    \label{fig:map_plot_SNR}
\end{figure}

\begin{figure}
    \centering
    \includegraphics[width=\linewidth]{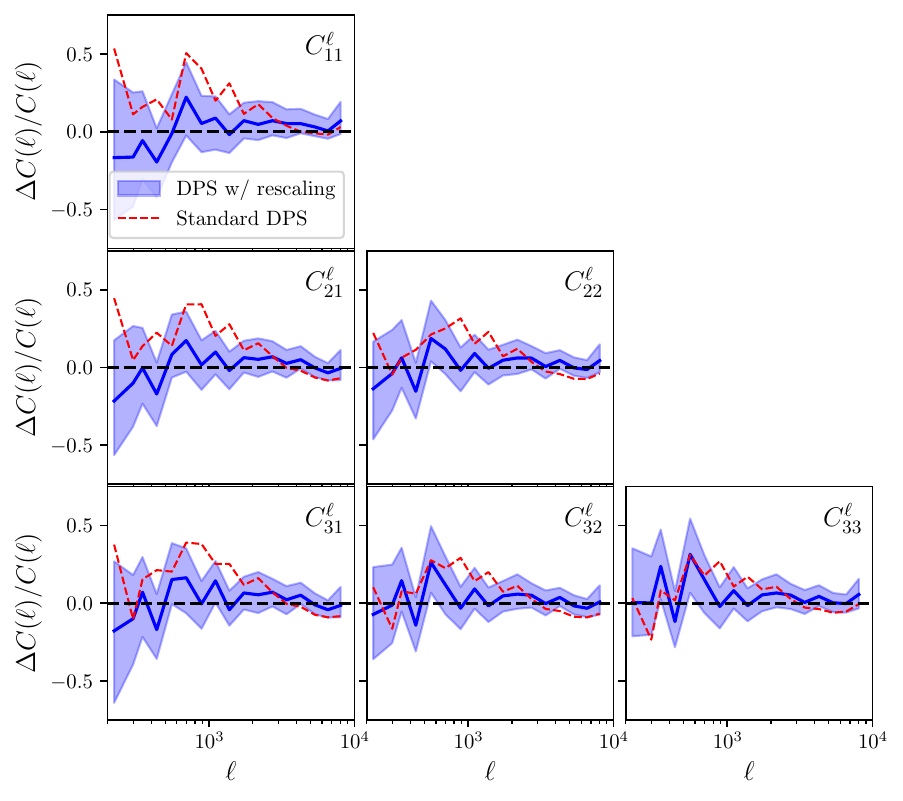}
    \caption{Fractional error in the power spectrum of the posterior maps with respect to the true underlying power spectrum of the simulation. We see that while the median of the standard DPS samples (red-dashed line) is highly biased. This bias is  mitigated by the likelihood rescaling method (shown in blue). Note that the power spectra shown are from a single sample of one patch. }
    \label{fig:posterior_Cl}
\end{figure}

\subsection{Mass map reconstruction with the diffusion model}\label{ssec:dps_mass_mapping}

Having validated our diffusion model, we now use our diffusion model to reconstruct weak lensing mass maps from the noisy data using the DPS algorithm. We utilize the method discussed in section \ref{ssec:dps_methodology} to sample posterior maps using our diffusion model.

\begin{figure*}
    \centering
    \includegraphics[width=\linewidth]{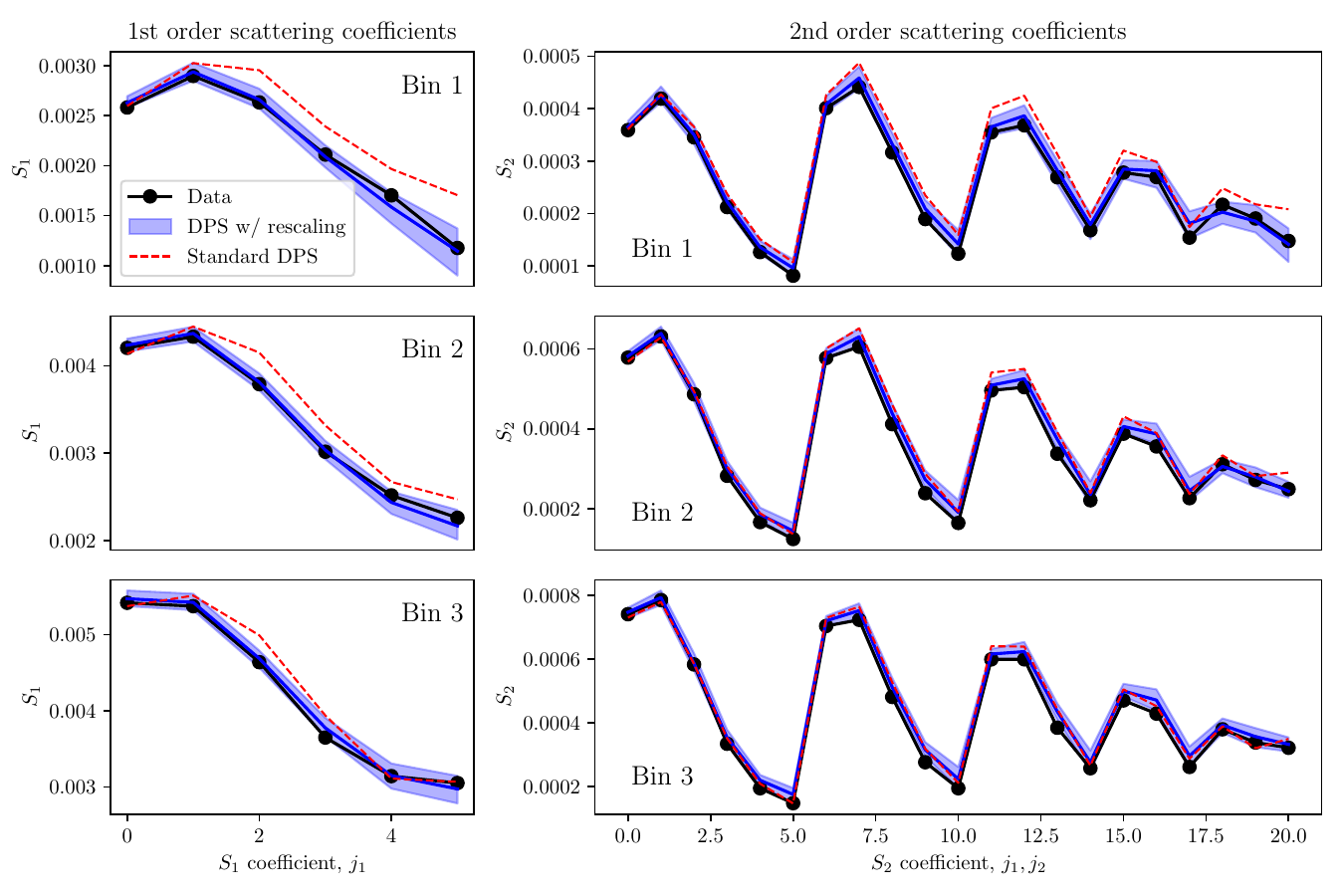}
    \caption{The first-order (\textit{left}) and second-order (\textit{right}) scattering transforms of the posterior maps. 
    The blue curves show these statistics for the DPS samples with the rescaling, red-dashed curves show the same statistics for the standard DPS method and the black points are the curves for the true underlying simulation.
    The three rows show the statistics for the three tomographic bins. As we can see from the figure, the DPS samples with the rescaled likelihood gives unbiased non-Gaussian summary statistics. }
    \label{fig:posterior_st}
\end{figure*}

As mentioned in that section, the standard DPS method leads to biased inference. To understand the origin of this bias, we plot the first-order scattering transform of the latent variable maps, $\mvec{x}_t$, at different time steps in Figure \ref{fig:S1_ood}. It is evident that at early times the DPS samples are not representative of the prior for the latent space. The errors accumulated in these early time steps are compounded and result in biased samples at $t=0$ as seen in Figures \ref{fig:posterior_Cl} and \ref{fig:posterior_st}. However, the likelihood rescaling version of DPS, equation \eqref{eqn:lkl_rescaling}, can mitigate this bias as shown in the figure. The green curves do not show any bias in $S_1$. 

We therefore use the rescaled version of DPS as the default posterior sampling method in the rest of this paper. We use it to infer the underlying mass maps from noisy shear data. The procedure to create our mock noisy data is outlined in section \ref{sssec:noisy_mock_data}. We sample a total of $100$ posterior maps. Computation for one sample takes $\mathcal{O}(50$ seconds$)$ on a NVIDIA TESLA-V100 GPU.   

Figure \ref{fig:map_plot_posterior} shows the convergence maps sampled using the DPS method. We show three randomly selected samples in the three middle panels. The mean of the $100$ sampled maps is shown in the second from the right panel.  From the figure, we see the the agreement between the true underlying map and the mean map and the posterior samples. 
The main advantage of a probabilistic mass mapping algorithm is that we can also quantify the uncertainty associated with the mass map reconstruction. This is shown in the right most panel of the Figure where we plot the standard deviation computed from $100$ posterior samples. As we can see from the figure, the DPS method gives us a way to characterize the full spatially correlated uncertainty in the mass map reconstruction.
In Figure \ref{fig:map_plot_SNR}, we plot the mean DPS map of the third tomographic bin and overplot the contours of the signal-to-noise ratio (SNR) of the maps defined as, 
\begin{equation}\label{eqn:kappa_snr}
    \text{SNR} = \frac{|\text{Mean}[\kappa_{\text{samples}}]|}{\text{Std. dev.}[\kappa_{\text{samples}}]}. 
\end{equation}
Here the mean and the standard deviation are computed from the sampled maps. The SNR $> 3$ peaks (voids) are highlighted with blue (red) contour lines showing that these peaks / voids will be robustly detected in the DPS mass maps from LSST-Y10 number densities. 

If the sampled mass maps are  fairly sampled from the posterior distribution, we expect the summary statistics of these maps to be unbiased. 
As such, we compare the power spectrum of the maps in Figure \ref{fig:posterior_Cl}. In that figure, we plot the fractional error in the power spectrum of the posterior maps with the blue curves. We do not see any significant biases in the posterior power spectra. Although we do note that there are residual biases of a few percent with our method. We can see from the figure that the power spectrum of the maps inferred using the standard DPS algorithm (i.e, without the likelihood rescaling) has a much larger bias which is mitigated by the likelihood rescaling approach. 

In Figure \ref{fig:posterior_st}, we plot the scattering transforms of the posterior maps. We compare the first order and second order scattering transforms. We see that the samples from the DPS posterior reproduces the scattering transforms of the true underlying maps showing that our method correctly captures the non-Gaussian information in the posterior maps. Similar to Figures \ref{fig:posterior_Cl}, we see a large bias in the scattering transform of the samples from the standard DPS algorithm that is mitigated with the likelihood rescaling. 
We show the performance of other non-Gaussian statistics such as the 1-point PDF, peak counts and void counts in Appendix \ref{app:ng_stats}.

Finally, we check the fidelity of the reconstructed maps by computing the pixel-by-pixel root mean square error (RMSE) and the Pearson correlation coefficient, $\rho_c$, defined as, 
\begin{align}
    \text{RMSE} &= \sqrt{\frac{1}{N_{\text{pix}}}\sum_{i=1}^{N_{\text{pix}}} (\kappa^i_{\text{rec}} - \kappa^i_{\text{truth}})^2}, \\
    \rho_{c} &= \frac{\langle \kappa_{\text{rec}} \kappa_{\text{truth}}\rangle}{\langle \kappa^2_{\text{rec}} \rangle \langle \kappa^2_{\text{truth}} \rangle},
\end{align}
where, $\kappa_{\text{truth}}$ is the true convergence map and $\kappa_{\text{rec}}$ is the reconstructed map. These metrics are commonly used for assessing the reconstruction quality \cite[e.g, ][]{DES_Y3_massmapping, Whitney2024}. 
The values for the RMSE and the correlation coefficients of the DPS maps are given in Table \ref{tbl:reconstruction_metrics} and compared with the corresponding numbers for a mass map reconstructed using the Kaiser-Squires (KS) inversion. The KS map is smoothed at 1 arcminute for optimal performance. We find that the DPS maps achieve superior performance over the KS maps, especially in the lowest redshift bin where the signal-to-noise is the lowest. For the first redshift bin, the DPS maps achieve $\sim 50\%$ improvement over the optimal KS map on both the RMSE and Pearson correlation coefficients.

\begin{table}
  \centering
  \caption{The RMSE and Pearson correlation coefficient, $\rho_c$, for the reconstructed DPS and KS maps. The KS map is smoothed at 1 arcminute for optimal performance. A smaller value of the RMSE and a larger value of $\rho_c$ corresponds to better reconstructions.}
  \begin{tabular}{l | c | c }
  \hline
      & RMSE & $\rho_c$ \\
      \hline
      DPS (Bin 1) & $0.014$ & $0.58$ \\
      KS (Bin 1) & $0.027$ & $0.38$ \\
      DPS (Bin 2) & $0.020$ & $0.62$ \\
      KS (Bin 2) & $0.027$ & $0.55$ \\
      DPS (Bin 3) & $0.026$ & $0.61$ \\
      KS (Bin 3) & $0.028$ & $0.61$ \\
    \hline
  \end{tabular}
  \label{tbl:reconstruction_metrics}
\end{table}

\section{Conclusion}\label{sec:conclusion}

\subsection{Summary}

Diffusion models have emerged as a powerful generative machine learning method for a range of scientific tasks. They have been used in multiple cosmological applications as a generative model for fast simulations and to reconstruct underlying cosmological fields from noisy data. These two tasks are often treated as separate: diffusion models trained for one purpose do not generalize to perform the other task. 

In this paper, we develop a single diffusion model that can be used for both tasks. By using the Diffusion Posterior Sampling (DPS) approach, we use a diffusion model trained to simulate weak lensing maps for the inverse problem of reconstructing mass maps from noisy weak lensing data. We find that the standard DPS method leads to biased inference but we  correct this bias by down weighting the likelihood term at early time steps of the diffusion. Our reconstructed mass maps have  the correct power spectrum and non-Gaussian summary  statistics (see Figs. \ref{fig:posterior_Cl},  \ref{fig:posterior_st} and \ref{fig:posterior_ng_stats}). The tests presented in these figures include tomography with three redshift bins and shape noise at the level of the LSST survey. While we have not included systematic errors, it would be straightforward to include simple models of intrinsic alignments and photo-$z$ in the forward model using additional channels or using conditional diffusion models. 

\subsection{Comparison to other mass-mapping approaches}

While several approaches to mass mapping have been developed since the pioneering work of Kaiser and Squires \cite{Kaiser1993}, we believe the fidelity and resolution of our mass maps is an improvement on most methods in the literature. We are not aware of any computationally feasible method for Stage 4 surveys that has been validated with the range of summary statistics presented here (see Figs. \ref{fig:posterior_st} and \ref{fig:posterior_ng_stats}).

Bayesian mass mapping methods utilizing Gaussian \cite{Loureiro2023, Sellentin2023} or lognormal \cite{Fiedorowicz2022, Fiedorowicz2022a, Boruah2024_massmapping} priors have been used to produce samples of probabilistic mass maps. However, both Gaussian and lognormal priors break down on small scales necessitating the use of alternate methods to reproduce the statistics of mass maps on small scales. One possible analytical extension is the use of Generalized Point Transformed Gaussian (GPTG) priors proposed in \cite{Zhong2024}. 

In the absence of a sufficiently accurate analytical model, ML-based methods are necessary for sampling mass maps from the posterior. A direct approach is training a generative model to sample mass maps from the posterior given noisy data \cite[e.g.][]{Floss2024, Whitney2024}. Another approach, which we adopt here, is to learn an ML-based prior. This prior can also serve as a generative model for the underlying field and then be integrated into a Markov Chain Monte Carlo (MCMC)-based sampling scheme \cite[e.g, ][]{Doeser2024}. However, MCMC-based methods suffer from long auto-correlation lengths and struggle with multi-modal posteriors \cite{Feng2018}, making them less scalable and computationally inefficient. Diffusion-based posterior sampling (DPS) addresses these issues by improving scalability and enabling efficient sampling from multi-modal posteriors.

To the best of our knowledge, \cite{Remy2023, Floss2024} are the only other examples of score-based mass mapping in the literature. But our strategy for mass mapping is significantly different from both these approaches. \cite{Floss2024} trained a score-based model to directly sample CMB lensing mass maps from the posterior. \cite{Remy2023} used an approach similar to ours of training the model to learn the generative model. However, their sampling strategy differs from ours -- they use HMC to sample their mass maps from the noisy posterior in the latent space and then used the reverse diffusion process to project these samples to get the posterior samples. 
We perform a more rigorous validation of our map reconstruction and show that the non-Gaussian features of our maps are well-reconstructed via comparison with a suite of summary statistics.  

In this study we have not explored the cosmological inference aspect since that would require sampling over cosmological parameters as well. We leave for future work the possibility of extending an approach like ours for field level inference. 


\subsection{Applications of the Diffusion model}

The diffusion based approach is computationally cheaper by a significant factor compared to direct simulations.  After the initial training, it takes only $\mathcal{O}$(10 seconds) to produce a single convergence map on a NVIDIA TESLA-V100 GPU. In contrast producing convergence maps without the generative models require running $N$-body simulations and then ray-tracing these simulations which may require upwards of a week to run. Thus the diffusion model represents many orders of magnitude speed-up in producing the simulated convergence mass maps. It is also much easier to implement in practice than other generative machine learning methods. We obtained reliable results with an off-the-shelf diffusion model. We have presented (Fig. \ref{fig:cov}) a first look at the accuracy of the covariance matrices estimated from our mass maps. The results are promising and open up the possibility of augmenting the available simulations by diffusion based synthetic maps to improve on the accuracy of covariance matrices especially for Higher Order Statistics (HOS) which are very demanding of simulation sample size. 

Lensing mass maps are a valuable data product with applications in cosmological inference and a range of studies of the large-scale structure in the Universe. The positive features of our diffusion based maps open up several applications such as in generating samples of  filaments, voids and clusters from noisy  lensing shear data. 

Large samples of galaxy clusters are a powerful cosmological probe.  
Several approaches that rely on cluster samples detected via the light (e.g. as red sequence galaxy overdensities) have proven to have subtle but problematic biases. 
These biases can be mitigated either by combining cluster abundance with clustering measures or by using SZ or X-ray peaks to define the cluster catalog. 
However the ideal cluster sample would be one detected from a mass map since the cluster observable-halo mass relation is bypassed \cite[e.g, ][]{Oguri2024}. 
The two limitations in current data are that mass maps are projected (solvable via forward modeling) and that the lower signal to noise enables only the most massive clusters to be detected. The mass mapping technique presented here can be used to improve on the second limitation by capturing all the information in the noisy shear data. 

The study of hot gas and galaxies as a function of type in cosmic filaments and voids is an active area of research in LSS \cite[e.g, ][]{Yang2020}. Again, the ideal filament/void detection approach relies on the lensing signal as the relationship of the tracer (gas or galaxies) to the mass can be directly studied \cite{Clampitt2016, Epps2017, HyeongHan2024}. Voids, the largest underdensities in the universe, are especially valuable tracers as their evolution is less impacted by mergers, anisotropies and other nonlinear effects \cite{Kovacs2022}. Our mass maps can be used to identify filaments and voids in survey data.


\begin{acknowledgments}
We thank Gary Bernstein, Marco Gatti, Matt Ho, Mike Jarvis, Laurence Perreault-Levasseur and Kunhao Zhong for useful discussions. S.S.B and B.J. are partially supported by the US
Department of Energy grant DE-SC0007901 and by NASA funds for the Open Universe project. 
\end{acknowledgments}

\bibliography{diffusion_posterior_sampling}%

\appendix
\section{Non-Gaussian statistics of the DPS maps}\label{app:ng_stats}

\begin{figure*}
    \centering
    \includegraphics[width=\linewidth]{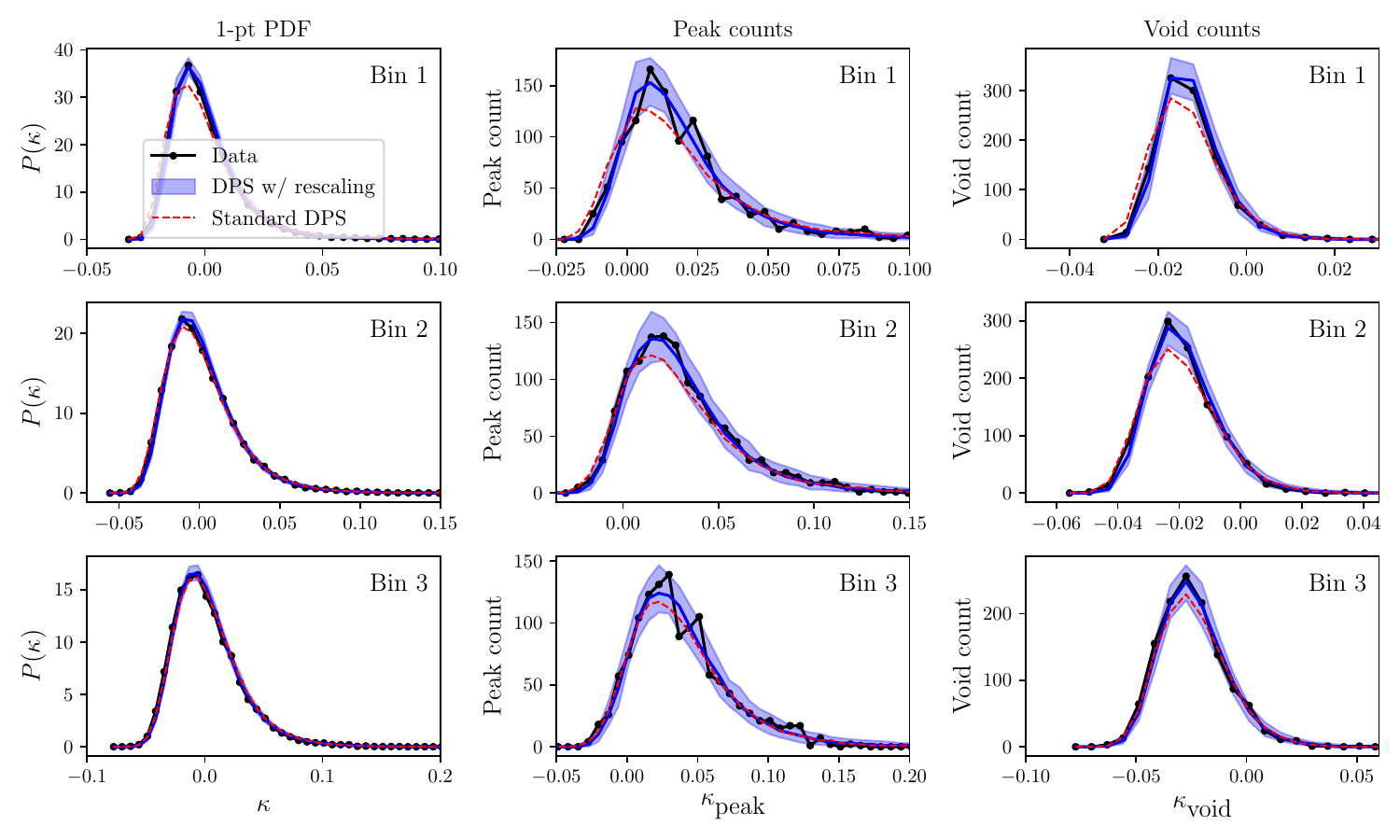}
    \caption{1-pt PDF (\textit{left}), peak counts (\textit{center}) and void counts (\textit{right}) of the posterior maps (\textit{in blue}) and compared to the same statistics of the true underlying map (\textit{black}). The median of the standard DPS samples are shown with the red dashed line. We can see that the standard DPS method typically underestimates the true distributions around the peak while our reconstructed maps give excellent agreement in all cases. }
    \label{fig:posterior_ng_stats}
\end{figure*}

To compare the quality of non-Gaussian reconstruction in our DPS maps, we compared the 1-point PDF, peak counts and void counts of the reconstructed maps in figure \ref{fig:posterior_ng_stats}. The statistics of the  posterior maps are shown in blue and compared to the same summary statistics of the true underlying map (in black). We see that the standard DPS method (red-dashed line) underestimate these statistics near the peak of the distribution.  

\end{document}